# The main features of the streamer discharge mode in coordinate gas detector

V.I.Razin

Institute for Nuclear Researches, Russian Academy of Sciences, Moscow

**Abstract.**

In work the results of the analysis are presented relating to the status of a streamer discharge mode in different gas detectors developed at the moment in laboratories and nuclear-physics installations. Thanks to new representations from a scope of high-speed narrow gas detectors together with the settle technique of a multiwire devices there are the possibility to undertake the next attempts to know a nature of this phenomenon. Studying prebreakdown and postbreakdown states in such detectors with the objective to rise their operation reliability under conditions of a high counting rate is a pressing issue.

*Introduction.*

Since opening by G.Charpak and F.Sauli [1] of a strong current operating mode of multiwire proportional chambers which later became know as self-quenched streamer (SQS) mode[2], passed more than 40 years. Nevertheless the nature of this physical phenomenon in gases at normal pressure has a number few the studied features. If before SQS-mode associated with large gap chambers, big percentage of extinguishing additives as hydrocarbons and methylal, thick anode wires etc. in modern wireless narrow –gap position-sensitivity devices of type RPC, GEM, MICROMEGAS [3-5], the existence of a large signal is also revealed.

The most typical feature of the transition of a gas-discharge detector to the streamer mode is a jump in the signal amplitude. In the transition region simultaneously small and large signals caused by the avalanche amplification mechanism and the streamer formation respectively exist. Signals with intermediate amplitudes are absent practically.This means that, as soon as the relevant conditions are created and streamer begins forming, its further development is a steady-state process that can continue even after the removal of the operating voltage from detector [6].

In the proportional gas amplification mode, in additional to the Townsend avalanche, others sources for increasing the ionization density in a gas may be considered (e.g., photoionization, recombination, metastable and nonmetastable Penning effect, association and dissociation reactions), which has affect the streamer onset. When analyzing the main streamer-development models it need to emphasize, that they give only a qualitative, illustrative description of this phenomenon but not a comprehensive explanation of the nature of the streamer mode.

Basic properties of the streamer mode have been revealed also in Micro-Pattern Gas Detectors (MPGD) at a high level of alpha-particle background [7]. Preliminary results of the research of the electron multiplication in MPGD expand borders of understanding of the streamer mode phenomenon. It is seen as a complex of electrostatic and electromagnetic interactions which begin with appearance of the precursor in plasma state. In an interelectrode gap, the plasma oscillations occur, accompanied by a longitudinal elastic waves of ionization, which can reach the cathode surface with induced negative charge. With the release of this charge due to previously established conducting channel , there is a strong current pulse, together with the emission due to recombination of positive and negative ions .As a result derive a thin cord or streamer.

The aim of this work is to consider and analyze the most typical features of the streamer mode. It may result in a deeper understanding of the streamer discharge nature. Consequently, studying  prebreakdown and postbreakdown phenomena in such detectors with the objective to raise their operation reliability under conditions of a high counting rate is a pressing issue.

*The main features of the streamer discharge mode.*

Thanks to new data it become possible to classify the main features of the streamer discharge mode, the presence of which is confirmed in most coordinate gas detector devices for the whole period of studies in this area in the following order:

1. The most typical feature of the transition of a gas-discharge detector to the streamer mode is a jump in a pulse amplitude. In a transit region simultaneously small and large signals caused by the avalanche amplification mechanism and the streamer formation, respectively exist. It should be noted also that there may be several disjoint curves at amplitude characteristic pertaining to the streamer mode.

2. There is a principal difference in the form of curve corresponding to the exponential dependence of the avalanche multiplication with the transition to saturation, and directly proportional to the increase of the streamer mode charge versus of the applied voltage.

3. Necessary but not sufficient condition for the avalanche to streamer transit is to achieve the Raether's threshold [8]:

$$\alpha \cdot N_0 \geq 10^8 \text{, where}$$

$\alpha$ -coefficient of amplification in avalanche,

$N_0$ –number of initial electrons

At the moment it knows that a probability of the streamer appearance has a maximum value in the gas mixtures in a comparison with a solely gas filling.

4. The plasma states with its characteristic features as slow electrons, electroneitralization of charges, electrostatic oscillations of the ion branch, etc. are more likely in the process of the avalanche multiplication with a gas gain of about $10^7$ -$10^8$ electrons [9].

5. Localization of streamers occur mainly along the drift axis of the primary electrons and the initial avalanches in the areas with a high content of excite and metastable atoms (dimers) of working gas at negligible small participation of the photoionization at this process.

6. Signal of the streamer mode is shaped like signals received in a pulse ionization chamber.

7. Streamer formation process is always associated with the appearance of pulse-precursor, which has a sufficiently large amplitude compared to conventional pulse amplitude of proportional mode and apparently associated with the development of a secondary avalanche in small field of a gas detector, where the positive space charge are existing with a maximum density.

In other words, in SQS –mode, there are three types of pulses:

a) proportional
b) avalanche precursor
c) streamer
The presence of these three components can be regarded as a sufficient condition for the occurrence of the streamer.

8. Plasma state in the gas volume of the detector can be create as a narrow luminous channel or streamer by passing of the complex electrostatic and electromagnetic interactions under release of induced negative charge from the cathode [9]

9. Plasma state can be realized in the extremely thin-gap gas discharge device with a width order 20 μm. In this case the electron must gain the energy for direct ionization process at electric field order 100 kV/cm according to equality $10^6 \equiv 2^{20}$.

10. Counting efficiency of the streamer events in relation to the full number of the pulses as a streamer and proportional modes is saturated and does not exceed a value of 50% after achieving of the certain threshold voltage between the electrodes of the detector.

11. Time delay in the appearance of the streamer pulse regard to the time of the pulse precursor originating can be from a few nanoseconds to hundred microseconds independently from the type of gas discharge coordinate devices [10]

Conclusion.

A brief analysis of the accumulated results related to streamer discharge mode shows that it is now seems possible to determine the necessary and sufficient conditions for the formation of the streamer in gas coordinate detectors. To use of the well-known laws from plasma physics to explain that phenomena it should be to consider the whole chain from the availability of the pulse-precursor which has large amplitude compared with the pulses of a proportional mode.

In same time should be studies its role in the time evolution of the second avalanches in the area with local maximum density of the positive space charge. It should be also take into account the effect of the presence of slow electrons with the high local electric field intensity equal to the external electric field, when the closure of the gap between electrodes is due to the passing of the elastic ionization wave. From this position can also be explained the fact of the sudden change (jump) of the current pulse after flowing of induce negative charge from a cathode to anode. After that a narrow glowing by recombination channel or streamer is appeared.

Great stimulation to study the nature of the streamer mode can be get by widely application now of the Micro-Pattern Gas Detector (MPGD) systems due to their excellent position resolution and capability to work at high counting rates. It is well know that devices of this kind are poorly defined of photon feedback mechanism, which can be to influence on occurring and propagation of the streamer. Another simplifying factor is that in the absence of wires the electric field is practically homogeneous, convenient to calculate together with basic localization discharge principle which declares: any discharge at the counter gas gap should produce only a local electric drop. So with a variety of specific properties MPGD as electron gas multiplication in the holes, the ability to step cascade avalanche development and to achieve a high gain without spark mode operation there are a good prospects to more qualitative studies of the streamer formation mechanism and the conditions of their transit to spark breakdown.

References

*1. G.Charpak and F.Sauli*, Nucl. Instr.and Meth. 96(1971),363.

*2. G.D.Alekseev at al.*, Nucl.Instr. and Meth. 177(1980),385.

*3. Y.Giomataris at al.,* Nucl.Instr. and Meth.in Phys.Res., A 306(1997),531.

*4 .Chechik R. at al.,* Nucl.Instr. and Meth.in Phys.Res., A 558(2006),475.

*5. Yu.N.Pestov,* Nucl.Instr.and Meth.in Phys.Res.., *A 494(2002),447-454.*

*6. V.I.Razin,* Instr.and Exper.Tech.,Vol.44,No.4,2001,pp.425-443.

*7. V.I.Razin and A.I.Reshetin*,Phys.of Part.and Nucl.Letters,2012,vol.9,№1,pp.58-61.

*8. H.Raether*,Electron avalanches and Breakdown in Gases, Washington,1964,pp.1-253.

 *9. Zalikhanov,B.Zh.,*Fiz.Elem.Chastits At. Yadra,1998,vol.29,no.5,p.321.

*10.P.Fonte, V.Peskov, B.Ramsey*, IEEE Trans. on N.S.Vol.46 ,No.3,June 1999,321.